\documentstyle[12pt]{article}
\begin{document}
\title{INTERSTELLAR HYDROGEN AND COSMIC BACKGROUND RADIATION}
\author{B.G. Sidharth$^*$\\ Centre for Applicable Mathematics \& Computer Sciences\\
B.M. Birla Science Centre, Hyderabad 500 063 (India)}
\date{}
\maketitle
\footnotetext{E-mail:birlasc@hd1.vsnl.net.in}
\begin{abstract}
It is shown that a collection of photons with nearly the same frequency
exhibits a "condensation" type of phenomenon corresponding to a peak
intensity. The observed cosmic background radiation can be explained from
this standpoint in terms of the radiation due to fluctuations in interstellar
Hydrogen.
\end{abstract}
In a previous communication\cite{r1} it was suggested that the origin of the
Cosmic Background Radiation is the random motion of interstellar Hydrogen. We
will now deduce the same result from a completely different point of view.\\
We start with the formula for the average occupation number for photons
of momentum $\vec k$ for all polarizations\cite{r2}:
\begin{equation}
\langle n_{\vec k} \rangle = \frac{2}{e^{\beta \hbar \omega}- 1}\label{e1}
\end{equation}
Let us specialize to a scenario in which all the photons have nearly the same energy so that we
can write,
\begin{equation}
\langle n_{\vec k} \rangle = \langle n_{\vec k'} \rangle \delta (k-k'),\label{e2}
\end{equation}
where $\langle n_{\vec k'} \rangle$ is given by (\ref{e1}), and $k \equiv
|\vec k|.$ The total number of photons $N$, in the volume $V$ being considered,
can be obtained in the usual way,
\begin{equation}
N = \frac{V[k]}{(2\pi)^3} \int^\infty_o dk4\pi k^2 \langle n_k \rangle\label{e3}
\end{equation}
where $V$ is large. Inserting (\ref{e2}) in (\ref{e3}) we get,
\begin{equation}
N = \frac{2V}{(2\pi)^3} 4\pi k'^2 [\epsilon^\theta - 1]^{-1}[k], \theta
\equiv \beta \hbar \omega,\label{e4}
\end{equation}
In the above, $[k] \equiv [L^{-1}]$ is a dimensionality constant, introduced to compensate the loss of
a factor $k$ in the integral (\ref{e3}), owing to the $\delta$-function in
(\ref{e2}): That is, a volume integral in $\vec k$ space is reduced to a
surface integral on the sphere $|\vec k| = k',$ due to our constraint that
all photons have nearly the same energy.\\
We observe that, $\theta = \hbar \omega/KT \approx 1,$ since by (\ref{e2}), the
photons have nearly the same energy $\hbar \omega$. We also introduce,
\begin{equation}
\upsilon = \frac{V}{N}, \lambda = \frac{2\pi c}{\omega} = \frac{2\pi}{k}
\quad \mbox{and}\quad z = \frac{\lambda^3}{\upsilon}\label{e5}
\end{equation}
$\lambda$ being the wave length of the radiation. We now have from (\ref{e4}),
using (\ref{e5}),
$$(e - 1) = \frac{\upsilon k'^2}{\pi^2}[k] = \frac{8\pi}{k'z}[k]$$
Using (\ref{e5})
we get:
\begin{equation}
z = \frac{8\pi}{k'(e-1)} = \frac{4\lambda}{(e-1)}[k]\label{e6}
\end{equation}
From (\ref{e6}) we conclude that, when
\begin{equation}
\lambda = \frac{e-1}{4} = 0.4[L]\label{e7}
\end{equation}
then,
\begin{equation}
z \approx 1\label{e8}
\end{equation}
or conversely.\\
Though equation (\ref{e7}) is quite general, as it stands, we have to assign
suitable units to it depending on the particular physical situation: We must get an additional input, in the order
of magnitude sense, from the system under consideration to fix the units.\\
Let us now consider the case of radiations due to fluctuations of the cold
interstellar Hydrogen as considered from an alternative view point in ref.\cite{r1}.
In this case it is infact known that
$\frac{1}{\upsilon} \sim 1$ molecule per c.c.\cite{r2,r3}. On the other hand,
the energy range for these cold molecules is small, so that the above
considerations apply. So from (\ref{e7}), owing
to the fact that $\upsilon^{1/3} \sim \lambda cm.,$ it follows that,
\begin{equation}
\lambda = 0.4cm.\label{e9}
\end{equation}
Remembering that from (\ref{e5}), $\lambda$ is the wave length and $\upsilon$ is the average volume per photon, the condition
(\ref{e8}) implies that all the photons are very densely packed as in the
case of Bose condensation. This means that from (\ref{e9}), we conclude
that at the wave length $0.4 cm,$ in the micro-wave region, the radiation
has a peak intensity. It is remarkable that the cosmic background radiation has the
maximum intensity exactly at the wave length given by (\ref{e9})\cite{r4}. So,
even without the Big Bang event it is possible to receive the observed
cosmic background radiation due to fluctuations in interstellar Hydrogen
(cf.ref.\cite{r3} and\cite{r5} for Hoyle and Wickramasinghe's attempt to explain the
background radiation in terms of Helium synthesis, without invoking the
Big Bang.)\\
The same conslusion can be obtained from yet another argument also\cite{r6}.
Interestingly, El Naschie has studied the Cosmic Background Radiation from the
point of view of fractal quantum space time\cite{r7,r8}.

\end{document}